\documentclass[11pt,letterpaper]{article}

\pdfoutput=1

\usepackage[margin=1in]{geometry}

\usepackage{cjhebrew}

\usepackage{setspace,caption}
\usepackage{footmisc}

\newlength{\myfootnotesep}
\usepackage[
  bookmarks=true,
  bookmarksnumbered=true,
  bookmarksopen=true,
  pdfborder={0 0 0},
  breaklinks=true,
  colorlinks=true,
  linkcolor=black,
  citecolor=black,
  filecolor=black,
  urlcolor=black,
]{hyperref}
\usepackage[round]{natbib}

\usepackage{xcolor}
\xdefinecolor{darkred}{rgb}{0.95,0.35,0.35}

\usepackage[normalem]{ulem}

\usepackage{tikz}
\usetikzlibrary{decorations.pathreplacing}

\usepackage{amsmath, amsthm, amssymb, amstext, comment, graphicx}
\usepackage{amssymb}
\usepackage{multirow}

\usepackage[capitalise]{cleveref}
\newcommand{\crefpart}[2]{\cref{#1}(\labelcref{#1-#2})}

\usepackage{enumitem}
\newlist{parts}{enumerate}{1}
\Crefname{partsi}{Part}{Parts}
\setlist[parts,1]{label=\alph*.,ref=\alph*}

\usepackage{fnbreak}

\usepackage{subcaption}

\usepackage[boxed]{algorithm2e}
\SetAlFnt{\onehalfspacing}
\SetAlCapSty{}
\SetAlCapNameSty{onehalfspacing}
\SetAlCapSkip{1em}
\SetKw{KwBy}{by}

\newtheorem{theorem}{Theorem}

\theoremstyle{definition}

\Crefname{figure}{Figure}{Figures}
\Crefname{algocf}{Algorithm}{Algorithms}

\newcommand{\Mechina}{PMA}
\newcommand{\Mechinot}{PMAs}
\newcommand{\ofMechinot}{PMAs'}

\hypersetup{
  pdfauthor      = {Yannai A. Gonczarowski <yannai@gonch.name>, Lior Kovalio <lior.kovalio@gmail.com>, Noam Nisan <noam.nisan@gmail.com>, and Assaf Romm <assaf.romm@mail.huji.ac.il>},
  pdftitle       = {Matching for the Israeli "Mechinot" Gap-Year Programs: Handling Rich Diversity Requirements},
}

\title{Matching for the Israeli ``Mechinot'' Gap-Year Programs:\texorpdfstring{\\}{ }Handling Rich Diversity Requirements\thanks{\protect\onehalfspacing First Draft: April 2018. A preliminary version of this paper appeared as an abstract in \emph{Proceedings of the 20th ACM Conference on Economics and Computation} (EC 2019). Gonczarowski was supported by the Adams Fellowship Program of the Israel Academy of Sciences and Humanities.
The work of Nisan was supported by ISF grant 1435/14 administered by the Israeli Academy of Sciences and by Israel-USA Bi-national Science Foundation (BSF) grant number 2014389.
This project has received funding from the European Research Council (ERC) under the European Union's
Horizon 2020 research and innovation programme (grant agreement No 740282).
The work of Romm was supported by a Falk Institute grant, by ISF grant 1780/16, and by a grant from the United States - Israel Binational Science Foundation (BSF).
We thank Josh Angrist, Parag Pathak, Tayfun S{\"o}nmez, and Bumin Yenmez, as well as the participants of the 2018 Workshop on Mechanism Design for Social Good, of MATCH-UP~2019, and of the 2019 NBER Market Design Working Group Meeting, for insightful comments.
We thank the Joint Council of Pre-Military Academies, and personally Dani Zamir, Yosi Baruch, Tamar Zeira, and Yaara Schur, as well as all of the individual \Mechinot, for their trust in us and for their cooperation throughout the redesign process. We also thank Ronny Goldstein and Matan Zur, who played a crucial role in the Joint Council of \Mechinot\ in successfully running the matching system in subsequent years.
}}

\author{Yannai A. Gonczarowski\thanks{\protect\onehalfspacing Microsoft Research. Research carried out while at The Hebrew University of Jerusalem, Israel. \emph{E-mail}: \href{mailto:yannai@gonch.name}{yannai@gonch.name}.} \and Lior Kovalio\thanks{The Hebrew University of Jerusalem, Israel. \emph{E-mail}: \href{mailto:lior.kovalio@gmail.com}{lior.kovalio@gmail.com}.} \and Noam Nisan\thanks{The Hebrew University of Jerusalem, Israel. \emph{E-mail}: \href{mailto:noam.nisan@gmail.com}{noam.nisan@gmail.com}.} \and Assaf Romm\thanks{Stanford University and The Hebrew University of Jerusalem, Israel. \emph{E-mail}: \href{mailto:assaf.romm@mail.huji.ac.il}{assaf.romm@mail.huji.ac.il}.}}

\date{August 25, 2020}

\begin{document}
\maketitle

\begin{abstract}
We describe our experience with designing and running a matching market for the Israeli ``Mechinot'' gap-year programs.  The main conceptual challenge in the design of this market was the rich set of diversity considerations, which necessitated the development of an appropriate preference-specification language along with corresponding choice-function semantics, which we also theoretically analyze. Our contribution extends the existing toolbox for two-sided matching with soft constraints. This market was run for the first time in January 2018 and matched 1,607 candidates (out of a total of 3,120 candidates) to 35 different programs, has been run twice more since, and has been adopted by the Joint Council of the ``Mechinot'' gap-year programs for the foreseeable future.
\end{abstract}

\vfill

\section{Background}

Israeli youth typically graduate from high school at the approximate age of 18 and are then required to enlist in the military for about two and a half years 
(currently a bit less for women and a bit more for men).  There are several institutions that offer high-school graduates a ``gap year'' before starting their
military service, mostly focusing on some combination of educational and volunteering activities.  It turns out that a significant number of youths, especially from
higher socio-economic statuses, are interested in taking such a gap year.  In addition to the core educational and volunteering activities,
these gap years typically also help them become more mature and independent, increase their self-confidence and their 
ability to get along with their peers, and build up their character, strengthening them for the challenging military service that lies ahead.\footnote{The public opinion of these programs in Israel seems to be that these are first and foremost opportunities to contribute to society by volunteering for a long period of time, in many cases with quite underprivileged populations (this could be compared in some sense to Teach for America in the USA), and that the personal gain in building character etc.\ is a ``fortunate consequence for the altruistic participants'' rather than the main reason for attending these programs.}  It seems that the military also sees a benefit from
these gap years, and hence it allows the participants to postpone their mandatory military service until the end of the gap year.

This paper centers around one of the most significant types of institutions that offer such gap years, called in Hebrew ``Mechinot Kedam Tseva'iyot'' (\cjRL{m:kiynwot q:dam--.s:bA'iywot}), 
in short ``Mechinot'' (\cjRL{m:kiynwot}),
and in English ``Pre-Military Academies'' (where ``pre-military'' is used in a strictly chronological sense), which we will abbreviate as \Mechinot.  The \Mechinot, in general, focus more time on study than on volunteering,
with an emphasis on various issues related to Jewish thought and to Israeli society, where the study is ``for the sake of studying'': with no grades and no certificates.  
There are over 50 different \Mechinot, and these are very
heterogeneous: some are religious, some secular, and some mixed; some are co-ed and some (mostly the religious ones) are single-gender; they have different mixes of activities, 
different focuses of studies, different
philosophical and social approaches (from purely religious orthodox to very progressive), and, one may comfortably say, quite different political leanings. Each \Mechina\ is independent and is separately run and administered, but they voluntarily cooperate with each other through ``The Joint Council of Pre-Military Academies'' (a cooperation that can be considered quite remarkable given the extreme range of social and political views represented).\footnote{See \url{https://mechinot.org.il/en-us/}.}

\looseness=-1
In this paper, we describe our experience with designing and running a national matching market for \Mechina\ admissions. Our matching system was run for the first time in January 2018, and has been used ever since. Let us first describe how the admissions process to the \Mechinot\ worked before~2018.  During the fall of their senior year of high school, candidates start considering which
\Mechinot\ they may want to go to in the following year (as well as several other options, including other types of gap years and not taking a gap year at all).  The most critical
part of the admissions process is an actual visit of the candidate to the \Mechina, usually spending a weekend there, participating in activities, getting to know the place, and
getting to be known (and informally evaluated) by the \Mechina\ staff.  These visits go on more or less continuously during the months from October to January.  

The \Mechinot\ see this inflow of candidates and \emph{build a group} for the upcoming year:  each \Mechina\ has a target number of candidates that it wishes to accept for the upcoming year --- usually between 25 and 100 ---
which is determined by various constraints, such as physical space and educational agenda and, perhaps most significantly, by the number of 
``military-service deferments'' granted to that \Mechina\ by the military (since each candidate must defer his or her service by a year).  The considerations employed by the \Mechinot\
are rather complicated and mirror their educational mission.  On the individual level, the \Mechinot\ look for qualities such as willingness
to learn and volunteer, and leadership potential,\footnote{There is high demand by \Mechinot\ for candidates who fit this profile.} but they also value affinity to their own educational agenda and special character.  More interestingly, and central to this paper,
most \Mechinot\ view their educational agenda also on a societal level, and so see being a meeting place for of the wide spectrum of the Israeli society a central part of their mission.  
The \Mechinot\ then build their groups incrementally from October throughout January, accepting candidates that they desire as these candidates visit the \Mechina, at each
point in time taking into account not only the candidate himself or herself, but also how he or she ``fits into the group built so far''. This allows each \Mechina\ in an online\footnote{The word \emph{online} is used in this paper in the CS/OR sense of processing the input serially, possibly making decisions before the entire input is given/realized.} fashion to correct any current gender imbalance,
make sure not to create a concentration of too many candidates from a single high school or small town,\footnote{While most of the \ofMechinot\ considerations while ``building a group'' revolve around diversity and balance, this specific point --- avoiding too many candidates from a single school --- is in fact important to many \Mechinot\ mainly due to peer effects, i.e., to avoid pre-existing cliques within the group that they build.}   implicitly maintain a wide variety of
balances according to different social criteria that the \Mechina\ cares about, as well as employ various affirmative-action policies for various less or more well-defined parts of Israeli society.  

\looseness=-1
Hence, during the several months of this admissions process (as it ran before our involvement), what transpired was a distributed online process where candidates continuously
visit \Mechinot, and then the \Mechinot\ continuously accept or reject candidates.   This is a difficult distributed task and
indeed it had not been working well: since the \Mechinot\ build their group in an online fashion, they need to know
which of the candidates that they accepted so far actually intend to come. This has led the \Mechinot\ to give ``exploding
offers'' to the candidates, where a candidate often has to accept an offer within days or forfeit his or her place.  The
candidates, many of whom view acceptance to a \Mechina\ as crucial,
find it very difficult to reply to these exploding offers as these come before they have even visited many of
the other \Mechinot\ (indeed, it was not uncommon for an attractive candidate to receive such an exploding offer from the first \Mechina\ that they visited) so they do not yet even have a clear picture of where they would like to go.  These two conflicting constraints
naturally led to much pressure, requests for extensions, strategic timing decisions, and ex-post change of heart on all sides.  
This process has led to significant inefficiency of the outcome both for the candidates who may need
to reject an offer due to having previously accepted 
an inferior one, and for
 the \Mechinot\ who need to make acceptance decisions before they have even seen
all of the candidates.
To this one must add the psychological burden due to even a reasonably successful candidate (and let alone a less successful one) naturally getting many rejections and being under significant uncertainty for a very long time. All of these are amplified by the unfortunate fact that there currently is a significant 
shortage of slots and so a large number of candidates will not get accepted to any \Mechina. Altogether, one can understand the general
dissatisfaction with this original system.

\section{Designing a Preference Specification Language}

During the year 2017, the authors of this paper have approached the Joint Council of \Mechinot\ and 
suggested switching to a centralized computerized admissions system.  As it turned out that the \Mechinot\ were well aware of the difficulties
with the existing system, they were rather happy to do so.  However, it was critical that each \Mechina\ maintain full educational sovereignty, as well as maintain an individual relationship with each 
candidate that it may accept. So the admissions process for 2018
(starting in
the fall of 2017) was done using a computerized system that was designed, built, and operated by the authors.   This
centralized matching excluded the religious \Mechinot\ (that have somewhat different circumstances), as well as
a few small \Mechinot\ that cater to special segments of the population.\footnote{Some of these did joint the match in subsequent years, and cite the good reputation of our system as one of the reasons for making an effort to be able to participate.}  Some of the \Mechina\ institutions offer more than a single
``program'' differing in character or in geographic location, and these 
different programs were viewed as separate \Mechinot\ by the matching system.  Altogether the system handled 35 different such programs 
(administered by 24 independent institutions), with a total of 1,760 slots, over which 3,120 candidates competed.

\begin{figure}[t]
\centering
\includegraphics[width=0.98\textwidth]{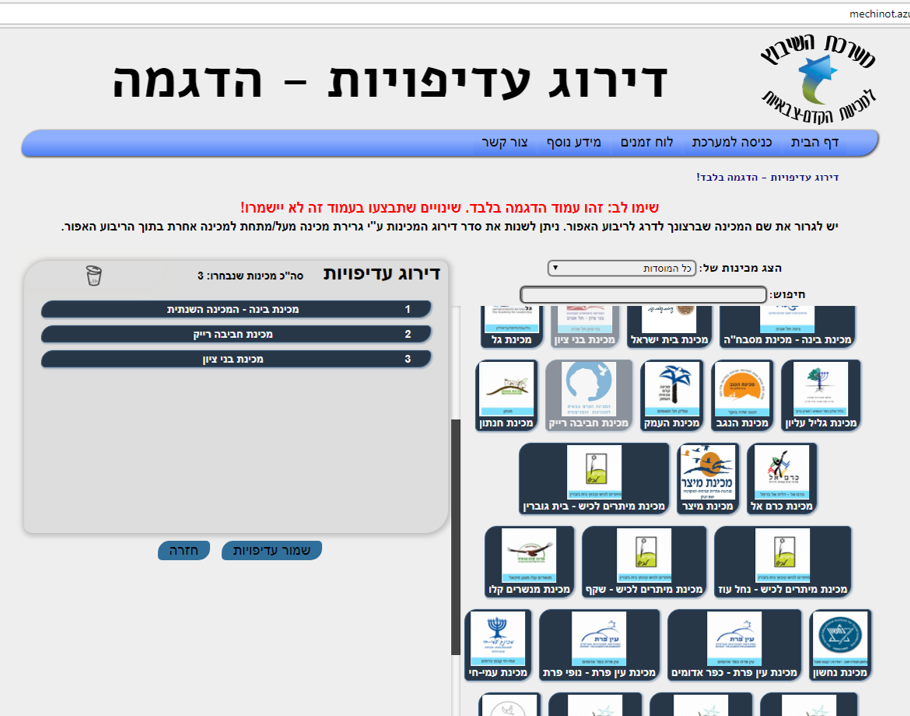}%
\caption{The graphical user interface by which a candidates specifies his or her preferences}\label{gui}
\end{figure}

\begin{figure}[t]
\let\translationfont\normalsize
\centering
\begin{subfigure}{\textwidth}%
\begin{tikzpicture}
\node at (0,0) {\includegraphics[width=.98\textwidth]{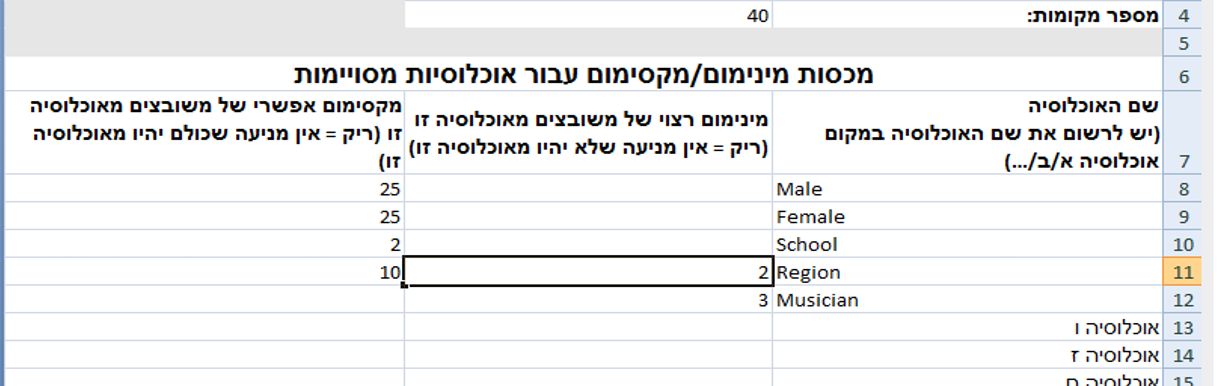}};
\node[darkred,font=\translationfont] at (-.75,2.35) {Number of Slots};
\node[darkred,font=\translationfont] at (4.7,-2.9) {Population};
\node[darkred,font=\translationfont] at (-.3,-2.9) {Minimum Target};
\node[darkred,font=\translationfont] at (-5.4,-2.9) {Maximum Quota};
\end{tikzpicture}%
\caption{Populations input sheet}%
\end{subfigure}
\vspace{1em}

\begin{subfigure}{\textwidth}
\begin{tikzpicture}
\node at (0,0) {\includegraphics[width=.98\textwidth]{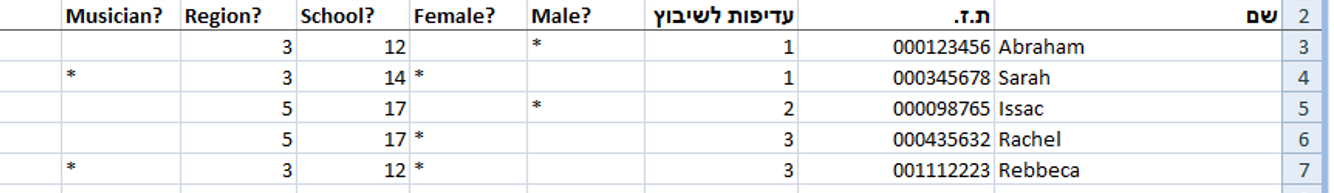}};
\draw [darkred,thick,decorate,decoration={brace,amplitude=10pt}] (-7.35,1.1) -- (-.25,1.1) node [midway,yshift=2em,align=center,font=\translationfont] {Titles generated according to \\ populations defined in sheet above};
\node[darkred,font=\translationfont] at (5.675,1.4) {Name};
\node[darkred,font=\translationfont] at (2.8,1.4) {ID};
\node[darkred,font=\translationfont] at (0.65,1.4) {Rankin\smash{g}};
\end{tikzpicture}%
\caption{Candidate input sheet}\label{excel-candidates}
\end{subfigure}
\caption{A spreadsheet by which a \Mechina\ describes its preferences (with titles translated to English)}\label{excel}
\end{figure}

The process 
worked as follows: from October 2017 through January 2018, the candidates visited the \Mechinot\ and ``interviewed'' there
as before in a distributed online manner.  However, the \Mechinot\ did not accept or reject any candidate during this
period, but rather only noted an evaluation of the candidate as well as any special attributes 
that they may care about when ``building a group'' (e.g., city of residence, gender, being religious, belonging to an underprivileged population or minority group, etc.). Similarly,
the candidates did not need to accept or reject any \Mechina, but simply noted to themselves how they evaluate each
\Mechina\ that they visited.   
During the first half of January~2018, each candidate had to login into a system that we deployed and enter his or her
ranking of
the visited \Mechinot.  Similarly, by mid-January, every \Mechina\ had to enter its preferences into the system using
a specially formatted Excel spreadsheet, which we describe below.  \cref{gui} shows a screenshot of the
graphical user interface that allows each candidate to create a ranking of \Mechinot\ by dragging and dropping their icons to form an
ordered list.\footnote{The candidates' graphical user interface was adapted from a similar system used for the Israeli Psychology Master's Match \citep{hrs2017}.}
\cref{excel} shows an example of a spreadsheet by which \Mechinot\ describe their preferences.

Once the preferences of
all of the candidates and \Mechinot\ were in the system, a variant of the Deferred Acceptance (DA) algorithm of \cite{gs62} was used to compute an assignment of candidates to
\Mechinot. This was motivated by the common view that stability (a property guaranteed by many variants of this algorithm --- see \cref{algorithm-barriers}) of the matching is essential for its success and survival \cite[see, e.g.,][]{roth2002} due to the independence of the \Mechinot. 
Furthermore, when preferences are ``well-behaved'' (for example, when institutions have responsive preferences), DA is strategyproof for the applicants \citep{df1981,roth1982}, and there was a strong desire to offer some strategic guarantees.

The main new ingredient that our system had to handle was the large set of ``diversity'' constraints that guide
the \Mechinot\ when ``building a group.''
There is a broad literature that deals with diversity and affirmative-action constraints, in particular through various types of lower and upper bounds, in two-sided matching markets.
One prominent line of work considers distributional constraints on institutions or on groups of institutions. This includes minimum quotas for specific institutions \cite[e.g.,][]{bfim2010,fituy2016} as well as quotas on groups of institutions \cite[e.g.,][and the references therein]{bfim2010,kk2015,kk2018}. In contrast, we are concerned not with inter-institution constraints on the number of applicants allowed in a set of institutions, but rather with intra-institution constraints on the number of applicants of a specific population within a specific institution. Such constraints in the form of minimum and maximum quotas were studied by \cite{huang2010} and subsequently \cite{fk2012}. Both of these papers consider a model with hard maximum and minimum quotas for every population in every institution. Such hard minimum quotas mean that invariably, in their model, no stable matching necessarily exists and so they focus on the study of whether a stable matching exists. Our problem differs from theirs since having had to actually build a matching mechanism, we had to output a matching even when no stable matching that perfectly matches all constraints exists. Considering affirmative action scenarios with two disjoint and complementing populations --- the ``majority students'' and the ``minority students,'' \cite{kojima2012} uses maximum quotas on the ``majority students'' population to effectively reserve seats for the ``minority students'' without explicitly using minimum quotas, thus avoiding such impossibilities. A similar approach to the two-population majority-minority model, which avoids some inefficiencies (such as unallocated slots) that manifest in this maximum-quotas approach, is taken by \cite{hyy2013}: their approach is for each school to give priority for any minority student until the school's ``minority reserve'' is filled, and beyond that to not prioritize based on populations. This approach was generalized by \cite{ehlers2014} and by \cite{ey2015} to any number of pairwise-disjoint populations. The requirements of the \Mechinot, though, involve not only far more than
two populations, but also heavily
intersecting populations. 
It indeed seems that the richness of the balance requirements that the \Mechinot\ desire does go significantly beyond previous analysis. A survey of the above and other approaches for handling ``quotas'' and diversity or affirmative-action-type constraints in stable matching markets can be found in a recent paper by \cite{nv17}.\footnote{\label{explainable}See also the very recent paper of \cite{abs2018} for an alternative approach based on Integer Programming. While the objective that their algorithm solves for is very clear and flexible, our algorithm enjoys incentive properties (see \cref{incentives} below) that were crucial in our case. Our algorithm is also somewhat more transparent in allowing to answer questions along the lines of ``why did candidate $c$ not end up at \Mechina\ $m$?'' --- questions that due to being managed independently of each other, many \Mechinot\ demanded answers to immediately following the match.}

Our approach is the following: devise a ``language'' that allows the \Mechinot\ to express their concerns when choosing among a group of candidates.  In our system, all quotas and all diversity and affirmative-action
constraints are expressed as part of the preferences of each of the \Mechinot\ separately (in the terminology of 
 \citealp{nv17} this is the ``modifying priorities'' approach).
As is common for such ``bidding languages'' \cite[see, e.g.,][]{n06}, the language must strike a compromise between different concerns: being expressive enough
as to (approximately) capture the real preferences of the \Mechinot\ on the one hand, and being simple enough to be handled well on the other hand.  This simplicity 
in our case has a triple meaning: algorithmic simplicity of running the match and obtaining an attractive outcome; strategic simplicity of handling the
incentives induced; and cognitive simplicity so that the humans running the \Mechinot\ can actually use it well.

\looseness=-1
Our language, encoded in the Excel spreadsheet format used by the \Mechinot\ depicted in \cref{excel}, is 
the following: each \Mechina\ defines a set of ``populations'' that it cares about (e.g., based on gender, religiousness,
city of origin, belonging to certain minority groups).  For each such population the \Mechina\ is allowed to define a \emph{maximum quota} as well as \emph{minimum target}.
In addition to ranking the candidates individually (a ranking that for cognitive simplicity may contain ties, which the \Mechinot\ were told would be broken randomly for them), the \Mechina\ also specifies set of populations to which each candidate belongs.\footnote{\label{multi-valued-populations}As can be seen in \cref{excel-candidates}, populations can be either ``binary'' (e.g., ``Male'' or ``Musician'') or ``multiple-valued'' (e.g., ``School'' or ``Region''). For the latter, the minimum target and maximum quota are applied to each ``value'' (e.g., specific school or region) separately. So, a multiple-valued population with, say, $10$ possible values, is no more than a shorthand for $10$ disjoint binary populations. The ability to have multiple-valued populations is an example of a feature which, despite not extending the expressiveness of the bidding language from a computer-science perspective, makes the language much easier to use for the humans who input the data (and is less error-prone in terms of, e.g., accidentally skipping a line), and in fact is a feature that we added following the request of several \Mechinot\ while the system was already operational. Starting in the following year, we went one step further and gave the \Mechinot\ the option of using auto-populated school and geographic populations. See \cref{subsequent}, which describes the changes that we made to our mechanism following its first year of operation, for more details.} We will henceforth refer to all of this information as a \emph{preference report}. Our interpretation of such a preference report
could be considered a generalization of the above-mentioned ``minority reserves'' approach of \cite{hyy2013} as well as of its above-surveyed extensions. We interpret a preference report as specifying a \emph{choice function} as follows: choose the top candidates according to their 
individual ranking, subject to the hard constraint that the set of candidates from any given population never exceeds
that population's maximum quota, and strictly preferring (in a way that overrides individual ranking) candidates that
belong to any population whose minimum target has not been reached.  There are several issues that need be specified
before this becomes fully formal, and these are spelled out in \cref{choice}\begin{algorithm}[t]
\SetKwInOut{Input}{input}
\SetKwInOut{Output}{output}
\Input{Set $C$ of candidates to choose from}
\Output{Set $\hat{C}$ of candidates chosen from $C$}
\tcp{Initialization}
$\hat{C}\leftarrow\emptyset$\tcp*{Chosen candidates}
\tcp{First pass:\ give higher priority to candidates who help reach minimum targets}
\ForEach{candidate $c\in C$, from highest-ranked to lowest-ranked}{
\If{$c$ is in some population whose minimum target is not met by $\hat{C}$}{
\If{adding $c$ to $\hat{C}$ does not violate any maximum quota}{
$\hat{C}\leftarrow\hat{C}\cup\{c\}$\;
}
}
}
\tcp{Second pass:\ all other candidates}
\ForEach{candidate $c\in C\setminus\hat{C}$, from highest-ranked to lowest-ranked}{
\If{adding $c$ to $\hat{C}$ does not violate any maximum quota}{
$\hat{C}\leftarrow\hat{C}\cup\{c\}$\;
}
}
\caption{The \emph{choice function} of a \Mechina, specifying for each set $C$ of candidates, which subset $\hat{C}\subseteq C$ of these candidates were to be chosen by the \Mechina\ (according to the preference report from its Excel spreadsheet) if that \Mechina\ had free choice from $C$ (and only from~$C$).  When traversing over candidates from highest-ranked to lowest-ranked, ties are broken randomly but consistently across \Mechinot\ (i.e., single tie-breaking, see \citealp{apr2009}).}\label{choice}
\end{algorithm}.
It is worthwhile to mention though that in terms of cognitive simplicity, virtually none of the \Mechinot\ 
showed any desire to dig into these issues beyond this (high level) general verbal interpretation, and were happy with this description as well as a separate high-level explanation of what stability means (see \cref{algorithm-barriers} below).

Of note in \cref{choice} is our design decision to give the same ``promotion'' in the choice function of a \Mechina\ to candidates that belong to any (positive) number of populations whose minimum target has not been reached, regardless of the (positive) number of such populations to which a candidate belongs. This decision was guided by the desire to prevent a \emph{systematic} phenomenon where each of only a handful of candidates ``fill up'' many affirmative action slots, which could have resulted in a matching system biased toward having as few candidates as possible benefit from such slots.
The very few \Mechinot\ that did inquire more about the details of the choice function were very content with the selection of choice function, and in particular from it potentially allowing more candidates to benefit from the priority slots that they allocated.
This design decision required some care, though: most \Mechinot\ had a separate pool of slots for men and for women; some of these \Mechinot\ initially expressed their gender-population preferences as, e.g., ``Male'' population with minimum target = maximum quota = 25, and ``Female'' population with minimum target = maximum quota = 25. The problem with expressing their gender-population preferences this way is that its effect is to completely ignore the minimum targets of \emph{all} populations, as any candidate who does not violate any maximum quota (and specifically, who does not violate their gender-population maximum quota) would have been given promotion since its gender minimum target would not have yet been met. For this reason, we advised these \Mechinot\ to drop the minimum targets that they set for each of the gender populations, leaving only the maximum quotas in place and resulting in proper promotion of candidates who help with ``real'' minimum targets.\footnote{\label{malefemale}Starting in the matching of the following year, we tweaked \cref{choice} to distinguish gender populations from all other populations, so that specifically a candidate that only helps with a minimum target for a gender population is given less promotion than a candidate who helps with a minimum target for any other population, but more promotion than a candidate who helps with no minimum target. See \cref{subsequent}, which describes the changes that we made to our mechanism following its first year of operation, for more details.}

For an alternative choice function that \emph{does} more significantly promote candidates that belong to a higher number of populations whose minimum target has not been reached, see \cref{choice-prefer-more}\begin{algorithm}[t]
\SetKwInOut{Input}{input}
\SetKwInOut{Output}{output}
\Input{Set $C$ of candidates to choose from}
\Output{Set $\hat{C}$ of candidates chosen from $C$}
\tcp{Initialization}
$\hat{C}\leftarrow\emptyset$\tcp*{Chosen candidates}
$\hat{R}\leftarrow\emptyset$\tcp*{Rejected candidates}
\While{$\hat{C}\cup\hat{R} \ne C$}{
\ForEach{candidate $c\in C\setminus\{\hat{C}\cup\hat{R}\}$}{
$n_c\leftarrow$ the number of populations of $c$ whose minimum target is not met by $\hat{C}$\;
}
$c\leftarrow$ the highest-ranked candidate in the set $\arg\max_{c\in C\setminus\{\hat{C}\cup\hat{R}\}}\{q_c\}$\;
\uIf{adding $c$ to $\hat{C}$ does not violate any maximum quota}{
$\hat{C}\leftarrow\hat{C}\cup\{c\}$\;
}\Else{
$\hat{R}\leftarrow\hat{R}\cup\{c\}$\;
}
}
\caption{An alternative choice function for a \Mechina, which gives higher priority to candidates that belong to a higher number of populations whose minimum target has not been reached. (In the interest of readability, we avoided any computational speedups such as updating the values of $n_c$ in every iteration of the while loop instead of recalculating them, keeping the candidates sorted and updating their order whenever any $n_c$ value changes, etc.)}\label{choice-prefer-more}
\end{algorithm}, which we also analyze below. We note that the decision to use \cref{choice} (rather than \cref{choice-prefer-more} or any other procedure) to give semantic meaning, as a choice function, to a preference report expressed using our spreadsheet-based preference-specification language is more than a normative decision on how to distribute priorities. Indeed, this decision restricts the choice functions that are describable using our preference-specification language to a certain space. As will be seen below, we evaluate our matching algorithm, both theoretically and empirically, by examining its output through the lens of the choice function of each \Mechina, and this evaluation is therefore only ever meaningful if the \emph{true} choice function of each \Mechina\ is close to the choice function that we have reconstructed for it from its preference report (using \cref{choice}). Adopting an inappropriate space of choice functions describable using our preference-specification language would have invariably forced these two choice functions --- the true one and the reconstructed one, which is the ``projection'' of the true one into this space (i.e., the closest choice function within this space) --- to be far from each other. Indeed, the decision to use \cref{choice} is first and foremost a descriptive decision and not merely a normative one. The actual lack of attempts by \Mechinot\ to deviate from our match is indicative, in retrospect, of the conclusion of our analysis below --- that the matching that we output was very close to stable\footnote{We emphasize that holding fixed an input preference report, specified using our spreadsheet-based preference-specification language, for each \Mechina, the decision between using \cref{choice} or \cref{choice-prefer-more} (or any other procedure) to interpret each preference report as a choice function
changes
which matchings are stable, as a matching that is stable with respect to one set of choice functions may well be unstable with respect to another.} --- holding not only with respect to the reconstructed choice functions, but also with respect to the true ones.

\section{Matching Algorithm and Theoretical Barriers}\label{algorithm-barriers}

Let us briefly review our model. There is a finite set of candidates and a finite set of \Mechinot. Each candidate has a strict preference order over a subset of the \Mechinot, where all other \Mechinot\ are considered unacceptable by this candidate. Each \Mechina\ is endowed with a choice function that is induced by \cref{choice} from some preference report (i.e., from some ranking, population definitions, upper quotas, and minimum targets that are expressible in our spreadsheet-based preference-specification language as explained above). Per the standard general definition \citep{hm2005}, a \emph{matching} is an assignment of each candidate to at most one \Mechina. A matching is \emph{individually rational} if it only assigns a candidate to a \Mechina\ that she ranks, and if each \Mechina, when given choice from the set of all of the candidates assigned to it, chooses all of these candidates. A candidate $c$ and a \Mechina\ $m$ form a \emph{blocking pair} with respect to a given matching if $c$ prefers $m$ to the \Mechina\ to which $c$ is assigned, and if $m$, when given choice from the set containing all of the candidates assigned to it and in addition also $c$, chooses a set of candidates that includes $c$. A matching is \emph{stable} if it is individually rational and has no blocking pairs.\footnote{So indeed given input preference reports, whether a given matching is stable depends on our decision of which algorithm (e.g., \cref{choice} or \cref{choice-prefer-more}) to use to interpret each \Mechina's preference report as a choice function.}\enlargethispage{1.4mm}

As mentioned above, our matching algorithm was a variant of the DA algorithm of \cite{gs62} that uses \cref{choice} (as the objective is to have good properties, such as stability or approximate stability, with respect to choice functions induced by \cref{choice}) to interpret the \ofMechinot\ preference reports, and is described in \cref{da}\begin{algorithm}[t]
\SetKwInOut{Input}{input}
\SetKwInOut{Output}{output}
\Input{Set $\mathcal{M}$ of \Mechinot\ with preferences, set $\mathcal{C}$ of candidates with preferences}
\Output{A feasible matching of (a subset of) $\mathcal{C}$ to (slots of) $\mathcal{M}$}
\Repeat{no candidate was rejected in this round}{
\tcp{A single deferred-acceptance round}
\ForEach{candidate $c\in\mathcal{C}$}{
$c$ applies to the \Mechina\ that $c$ ranks highest of those who have not (yet) rejected $c$\;
}
\ForEach{\Mechina\ $m$}{
$C\leftarrow$ the set of candidates that applied to $m$ in this round\;
$\hat{C}\leftarrow$ the set of candidates chosen from $C$ via $m$'s choice function (\cref{choice})\;
$m$ rejects all candidates in $C\setminus\hat{C}$\;
}
}
\tcp{Match according to the last round above}
\ForEach{candidate $c\in\mathcal{C}$}{
Match $c$ with the \Mechina\ to which it applied in the last round above\;
}
\caption{Deferred-Acceptance (DA) using choice functions defined using \cref{choice}.}\label{da}
\end{algorithm}.
The preferences of the candidates (who are the proposing side in our implementation) fit exactly the scenario of \cite{gs62}: a simple preference order on (a subset of) the \Mechinot.
The preferences of the \Mechinot, however,
are significantly more complex: beyond a preference order on the candidates, they also have maximum quotas and/or minimum targets on various populations, all expressed through their choice functions. Notably, a \Mechina's choice function does not necessarily satisfy substitutability \citep{roth1985,hm2005},\footnote{Nor of course does it satisfy weak substitutability \citep{hk2008}, unilateral substitutability \citep{hk2010}, or substitutable completability \citep{hk2015}, as all of these coincide with substitutability for matching markets without contracts, such as the \Mechinot\ market.} nor can it be described using slot-specific priorities \citep{ks2016}.
Can the desirable theoretic guarantees that  DA has when preferences are ``well behaved'' be extended to this scenario as well?\footnote{Since (weak) substitutability is known to also be necessary, in the maximal domain sense, to guarantee the existence of a stable matching \citep{hk2008}, some further assumptions must of course be made to restore any of the above theoretical guarantees.}

As noted above, \cite{huang2010} studies similar constraints. However, in that paper the minimum constraints are hard and binding.
In the absence of (positive) minimum targets/quotas, though, our model and choice functions coincide with those of that paper. 
That paper focuses on whether or not a stable matching exists, and observes that this turns out to depend on the structure of overlap of the different populations. If the set of populations for each institution is \emph{laminar} (i.e., every two populations are either disjoint or one contains the other), then that paper shows that in the absence of minimum quotas (and so, equivalently, in the absence of minimum targets in our model), a stable matching always exists. Furthermore, that paper shows that in the presence of minimum quotas, while laminarity obviously cannot guarantee the existence of a stable matching, laminarity does give rise to a polynomial-time algorithm for \emph{deciding whether} a stable matching exists. In the absence of laminarity, as that paper shows, such a decision turns out to be NP-hard. As noted above, we consider ``soft'' minimum targets rather than ``hard'' minimum quotas as in that paper. We show, when choice functions are defined via \cref{choice}, that laminarity of the populations, in addition to a certain condition on the structure of those populations that also have minimum targets (and not just maximum quotas), guarantees that a stable matching nonetheless exists and can furthermore be efficiently found in a \emph{strategyproof} manner (the proof is starkly different from that of previous papers, though).\footnote{The crucial role of laminar populations is also identified in independent work by \cite{bs2019a,bs2019b} that studies the problem of constitutional implementation of reservation policies for affirmative action in India. They use the term ``nested'' for this population structure.}

\begin{theorem}\label{laminar-disjoint}
Let \cref{choice} be used to interpret the \ofMechinot\ preference reports as choice functions --- to be used both when running DA and when evaluating a matching for stability.
If the set of populations in the preference report of each of the \Mechinot\ (separately) is laminar,
and furthermore for each \Mechina\ the populations that have (positive) minimum targets are pairwise disjoint,
then DA produces a stable matching and is strategyproof for the candidates.
\end{theorem}

\cref{laminar-disjoint} can be seen as generalizing certain aspects of the ``minority reserves'' approach of \cite{hyy2013}, where there are only two disjoint populations, only one of which has a minimum target, and of its above-surveyed extensions that assume pairwise-disjointness of all populations. \cref{laminar-disjoint} and the other \lcnamecrefs{laminar-chains} stated in this section are proved in \cref{analysis}. \cref{laminar-disjoint} is in fact a corollary of a more general \lcnamecref{laminar-chains} that we prove, which shows that when interpreting preference reports using \cref{choice-prefer-more} (in lieu of \cref{choice}), a weaker condition on the structure of the populations that have minimum targets suffices.

\begin{theorem}\label{laminar-chains}
Let \cref{choice-prefer-more} (rather than \cref{choice}) be used to interpret the \ofMechinot\ preference reports as choice functions --- to be used both when running DA and when evaluating a matching for stability.
If the set of populations of each of the \Mechinot\ (separately) is laminar,
and furthermore for each \Mechina\ the set of populations that have (positive) minimum targets is a union of pairwise-disjoint chains,\footnote{That is, if a population $P$ has a (positive) minimum target and contains two populations $P'\subsetneq P$ and $P''\subsetneq P$ that are incomparable (i.e., $P'\not\subseteq P''$ and $P''\not\subseteq P'$), then $P'$ and $P''$ cannot both have (positive) minimum targets.}
then DA produces a stable  matching and is strategyproof for the candidates.
\end{theorem}

The key idea behind the proof of \cref{laminar-chains} is to show that the choice function defined by \cref{choice-prefer-more}, under the assumed population structure, is \emph{substitutable} and satisfies \emph{irrelevance of rejected contracts}\footnote{As we show, irrelevance of rejected contracts in fact holds even in the absence of any structural assumptions on the populations.} and the the \emph{law of aggregate demand}. With these three properties in hand, \cref{laminar-chains} follows from the analysis of \cite{hm2005} and \cite{as2013}.
Recall that in the absence of minimum targets/quotas, \cref{choice,choice-prefer-more} coincide with each other and with the choice function of \cite{huang2010}. Therefore, the beginning of our proof of \cref{laminar-chains}, which proves the \lcnamecref{laminar-chains} for this case, in fact constitutes not only a full concise alternative proof of the theorem of \cite{huang2010} that in the absence of any minimum quotas a stable matching exists,\footnote{In addition to the conciseness of our proof, it does not seem that \citeauthor{huang2010}'s techniques lend themselves to adaptation to our setting when minimum targets are present.} but also a proof that then any choice function is substitutable and satisfies irrelevance of rejected contracts and the law of aggregate demand. This also implies many additional desirable properties \citep{roth1984,hm2005,as2013} including that DA is strategyproof.\footnote{One of these additional properties is the existence of a lattice structure for the set of stable matchings. See \cite{fk2012} for a proof of this property in particular, under hard minimum quotas in a generalization of the model of \cite{huang2010}.}
To see that \cref{laminar-disjoint} indeed follows from \cref{laminar-chains}, observe that a pairwise-disjoint set is a special case of a union of pairwise-disjoint chains, and that  when all minimum-target populations are pairwise disjoint, \cref{choice,choice-prefer-more} coincide. As neither condition on the structure of the minimum-target populations is actually fully satisfied by the \ofMechinot\ real populations, as noted above we opted to use \cref{choice} to interpret the \ofMechinot\ preference reports despite its slightly worse theoretical properties demonstrated above, due to our belief that the choice functions reconstructed by it better represent the \ofMechinot\ real preferences.\footnote{This argument, which led us to choose to use \cref{choice} to interpret the preferences of the \Mechinot, is indeed somewhat less convincing in a context (\emph{unlike} ours) in which the set of minimum-target populations is in fact a union of pairwise-disjoint chains. Indeed, in such a context it is not possible to have two intersecting minimum-target populations $P$ and $P'$ (say, each with a minimum target of $1$) and two candidates $c$ and $d$ such that $c$ belongs to $P$ but not to $P'$, and $d$ belongs to $P'$ but not to $P$. Indeed, given such a population structure, if $c$ belongs to $P$ and $d$ belongs to $P'$, invariably one of $c$ and $d$ belongs to both $P$ and $P'$. (So, it is not the case that a single candidate $e$ that belongs to both $P$ and $P'$ and benefits from both minimum targets can be replaced by $c$ and $d$, allowing each to benefit from one of the minimum targets.)}

\looseness=-1
To prove that under the conditions of \cref{laminar-chains} (and hence of \cref{laminar-disjoint}) each choice function is substitutable and satisfies irrelevance of rejected contracts and the law of aggregate demand, we show that for every candidate set $C$ and for every $c\in C$, if we denote by $\hat{C}$ the set chosen from~$C$ by a given \Mechina\ and by $\hat{C}'$ the set chosen from $C\setminus\{c\}$ by the same \Mechina\ (with the same preferences), then if $c\notin\hat{C}$ then $\hat{C}'=\hat{C}$ (irrelevance of rejected contracts), and otherwise either $\hat{C}'=\hat{C}\setminus\{c\}$ or $\hat{C}'=(\hat{C}\setminus\{c\})\cup\{d\}$ for some $d\in C$. This in particular implies that $\hat{C}\setminus\{c\}\subseteq\hat{C}'$ (substitutability) and that $|\hat{C}'|\le|\hat{C}|$ (law of aggregate demand). To prove this, we trace the invocation of \cref{choice-prefer-more}, in parallel both for the input $C$ and for the input $C\setminus\{c\}$. If these two invocations of \cref{choice-prefer-more} do not have the same candidates chosen at the same steps, then the first difference must be for candidate $c$ to be considered and chosen in the former invocation, but not in the latter. This holds as at any point, the set of candidates chosen so far (along with the reported preferences) effectively defines an order on all not-yet-considered candidates, according to which the next candidate to be considered is chosen. This is true even in the absence of any structural assumptions on the populations. From that point onward, as long as the same candidates are chosen in both invocations and in the same steps, for any population the number of chosen candidates from this population in the latter invocation is less than or equal to the same number in the former invocation. Therefore, either both invocations accept the exact same candidates from that point, in which case $\hat{C}'=\hat{C}\setminus\{c\}$, or the chronologically next difference in chosen candidates between both invocations is that some candidate $d$ is chosen in the latter invocation but not in the corresponding step in the former invocation. In the case of no minimum targets, we can show that laminarity of the populations guarantees that there can be no more differences between the two invocations after that point (and so $\hat{C}'=(\hat{C}\setminus\{c\}\cup\{d\})$), hence concisely reproving the theorem of \cite{huang2010} that in the absence of any minimum quotas (hard quotas in the case of that paper, soft targets in our case) a stable matching exists. We furthermore we get ``as a byproduct'' strategyproofness of DA as well as all of the other desirable properties \citep{roth1984,hm2005,as2013} that were not proven by \cite{huang2010} but follow from substitutability, irrelevance of rejected contracts, and the law of aggregate demand. If there are minimum targets, then additional differences between the two invocations are in fact possible. Nonetheless, as we show, the structural properties of the minimum-target populations (i.e., union of pairwise-disjoint chains) still allows some careful bookkeeping to yield the desired result. Specifically, we show that the chronologically next difference, if any, in chosen candidates must be for $d$ to be chosen in the former invocation (so if there are no more differences, then $\hat{C}'=(\hat{C}\setminus\{c\})$); that then the chronologically next difference, if any, must be for some candidate $e$ to be chosen in the latter invocation but not in the former (so if there are no more differences, then $\hat{C}'=(\hat{C}\setminus\{c\})\cup\{e\}$); that then the chronologically next difference, if any, must be for $e$ to be chosen in the former invocation (so if there are no more differences, then $\hat{C}'=(\hat{C}\setminus\{c\})$); etc. (See \cref{analysis} for the full proof.)

Unfortunately, in our setting neither are the populations in any sense laminar as in \cref{laminar-disjoint} or \cref{laminar-chains}, nor do the minimum-target populations satisfy the respective structural properties. For example, a ``geographic population'' and a ``religion population''
typically neither are disjoint nor have one containing the other. As noted above, \cite{huang2010} has shown that even without any minimum targets/quotas, for nonlaminar populations it is possible that no stable matching exists. In this spirit, we show that dropping the condition on the structure of populations with minimum targets similarly destroys the desired theoretical properties (which in our case include both existence of a stable matching and strategyproofness):

\begin{theorem}\label{impossibility}
Let \cref{choice} be used to interpret the \ofMechinot\ preference reports as choice functions --- to be used both when running DA and when evaluating a matching for stability.
If each \Mechina\ has laminar populations, yet its minimum-target populations can intersect, then it is possible that no~stable matching exists, and DA is not strategyproof. Furthermore, both of these hold even if the set of minimum-target populations of each \Mechina\ is guaranteed to be a union of pairwise-disjoint chains.
\end{theorem}

\looseness=-1
The crux of the proof of \cref{impossibility} is to look, as an example, at a \Mechina\ $m$ that ranks three candidates $c,d,e$ in the order $c\succ d\succ e$, and considers three populations: $\{c,d,e\}$ with minimum target~$1$, $\{d,e\}$ with maximum quota $1$, and $\{e\}$ with minimum target $1$. For such a \Mechina, it is possible that $e$ be rejected in favor of $d$ during a run of DA (due to the maximum quota of~$1$ on the population comprised of both), but later in that run for $c$ to also be accepted by $m$, in which case \Mechina\ and $e$ block (since without $c$, candidate $d$ is preferred to candidate $e$, but with $c$, candidate $e$ is preferred to candidate $d$ since only the former satisfies a not-yet-satisfied minimum target). 
The possibility of a non-substitutable choice function, such as the one just demonstrated, in fact also motivated our decision to use single tie-breaking (rather than multiple tie-breaking, see \cref{choice}) when running DA. Indeed, it is not hard to see that a necessary condition for a blocking pair to form in DA due to choice functions not necessarily being substitutable is for a candidate to be rejected after he or she had already caused another candidate to be rejected (as in the above example). Single tie-breaking, by correlating a candidate's desirability across \Mechinot, reduces the probability of such an event.

\section{Dealing with General Preferences}\label{evaluate}

Despite \cref{impossibility}, based on our understanding of the problem domain, as well as on trial runs with various parameters, we believed that DA (using \cref{choice} to interpret the \ofMechinot\ preference reports) would give rather good results in practice, and this is what we set on to implement.  The big question was to what extent we could evaluate the quality of this algorithm and support
this belief. We now turn to discuss this, while paying special attention exactly to the properties that \cref{impossibility} shows we cannot fully achieve: stability and strategyproofness.

The level of strategyproofness is difficult to evaluate algorithmically, and an ex-post evaluation is also
less satisfying.  To nonetheless give a useful strategic guarantee, we were able to formally prove that the two main types of manipulation that our candidates were considering
are not profitable.
Informal conversations with candidates and with parents of candidates while we were designing the system and toward the deadline for submitting their preferences suggested that they only considered or worried about the following two
types of manipulation, each of which was brought up by quite a few candidates/parents:
\begin{itemize}
\item \emph{Truncation}: many candidates were worried that listing multiple \Mechinot\ in their preference list may cause the
system to match them to a less preferred option rather than a more preferred one that they would have gotten 
had they not allowed the system to assign them to the less preferred option.
\item \emph{Sure thing}: many candidates knew that they have a ``guaranteed slot'' in a certain \Mechina, i.e., they were assured that the \Mechina\ did not rank above them more candidates than
it has slots (for any population).  Such candidates were often worried they could lose the guaranteed slot \linebreak unless they ranked this certain \Mechina\ at the top even when they really preferred another \Mechina.
\end{itemize}
We prove formally that neither of these two manipulations can ever be profitable, which can be contrasted with known results for other mechanisms or for other sides of the market.\footnote{For a formal and systematic treatment of a conceptually similar (but technically incomparable) approach of ruling out ``obvious manipulations,'' see the recent paper by \cite{tm2020}.}
The proof identifies these two properties, and in particular the latter, as very robust properties at the core of ``candidate-proposing'' DA, and is readily generalizable to many other variants of this algorithm. Resilience to such manipulations is important also given the documented potential gains from simple manipulations in many popular mechanisms \citep{rr1999,mwps2015}.

\begin{theorem}\label{incentives}
Regardless of the population structure,
DA has the following incentive properties for every candidate and every
profile of preferences, regardless of whether \cref{choice} or \cref{choice-prefer-more} is used to interpret the \ofMechinot\ preference reports as choice functions:
\begin{parts}
\item\label{incentives-truncation}
If when specifying a truncated preference list, i.e., a preference list obtained by removing one of more \Mechinot\ from the end of this candidate's original preference list, the candidate is assigned to a certain \Mechina, then the candidate would have been assigned to the same \Mechina\ even if she submitted her original, non-truncated, preference list.
\item\label{incentives-sure-thing}
If the candidate is guaranteed to be assigned to a certain \Mechina\ if the candidate ranks it first, where ``guaranteed'' means 
that this would happen for any profile of preference lists of the other candidates,\footnote{Specifically, if the choice function of that \Mechina\ always chooses this candidate from any set of candidates that contains it.} then if the candidate specifies any preference list that contains this \Mechina, the candidate is still guaranteed to either be assigned to that \Mechina\ or to one that the candidate ranked higher on her preference list.
\end{parts}
\end{theorem}

Moving on from strategyproofness to stability,
to measure the extent of the lack of stability of a matching (once such is reached), we turned to the standard measure of \emph{blocking pairs}. Recall that a candidate $c$ and a \Mechina\ $m$ block an assignment if\ \ a)~$c$~prefers~$m$ over the \Mechina\ to which $c$ is assigned, and\ \ b)~if $m$ were to choose from the set of all candidates assigned to it and in addition $c$, then the choice of $m$ would include~$c$. (We also used other related measures such as the number of candidates involved in blocking pairs.)  In trial runs with various
natural distributions, we noticed that typically there are very few blocking pairs, which furthermore involve only a small number of \Mechinot, and within a \Mechina\ only a small number of populations.  This was also borne out on the real 
data --- see \cref{epilogue} as well as \cref{stats} in \cref{subsequent}.

One specific type of blocking pair that our trial runs (as well as the real data --- see \cref{epilogue} as well as \cref{stats} in \cref{subsequent}) showed that DA produces deserves special discussion: a blocking pair that can be resolved without harming any candidate. Such a blocking pair involves a \Mechina\ $m$ that has not fulfilled its quota and a candidate $c$ (who may be unmatched or matched to a \Mechina\ that he or she ranked below $m$) such that $c$ can be matched to $m$ without rejecting any candidate currently matched to $m$ (equivalently, the choice function of~$m$, when applied to the set that includes all candidates assigned to $m$ as well as $c$, chooses this entire set).\footnote{Such a blocking pair may form, for example, in a situation along the following lines: say that $m$ has capacity~$2$, has maximum quota $1$ for a population~$P$ comprised of candidates $c$ and $d$, and has maximum quota $1$ for a population $P'$ comprised of candidates $d$ and $e$. Then it is possible that $c$ is rejected by $m$ in favor of $d$ during a run of DA, but that later in that run $d$ is rejected by $m$ in favor of $e$. At that stage $c$ and $m$ are blocking. (Note that the population structure here is not laminar. Indeed the proof of \cref{laminar-disjoint,laminar-chains} shows that the conditions of these \lcnamecrefs{laminar-chains} on the population structure must be violated for such a blocking pair, or for any blocking pair, to form.)} Such a blocking pair can be easily resolved: simply transfer/assign $c$ to $m$. Since such a resolution of this blocking pair constitutes a Pareto improvement for all candidates, then iteratively resolving any such pairs until no more such pairs exist cannot cause an infinite loop. Observing this, we heuristically added a final stage to our algorithm: a candidate-Pareto-improvement stage described in \cref{improvement}\begin{algorithm}[t]
\SetKwInOut{Input}{input}
\SetKwInOut{Output}{output}
\Input{Set $\mathcal{M}$ of \Mechinot\ with preferences, set $\mathcal{C}$ of candidates with preferences}
\Output{A feasible matching of (a subset of) $\mathcal{C}$ to (slots of) $\mathcal{M}$}
\tcp{Stage 1:\ Deferred acceptance}
Compute a matching via DA (\cref{da})\;
\tcp{Stage 2:\ Candidate Pareto improvement}
\While{$\exists$ blocking pair $(m,c)$ s.t.\ $c$ can be matched to $m$ without rejecting any candidate}{
$m\leftarrow$ some \Mechina\ that is part of a blocking pair as defined above\;
$c\leftarrow$ the candidate ranked highest by $m$ s.t.\ $(m,c)$ is a blocking pair as defined above\;
Match $c$ with $m$ (breaking any previous match of $c$ with another \Mechina, if existed)\;
}
\caption{The algorithm that we ran, consisting of deferred acceptance (\cref{da}) and an added candidate-Pareto-improvement stage.}\label{improvement}
\end{algorithm}. While we were able to contrive an example where \crefpart{incentives}{truncation} breaks for \cref{improvement} due to the added candidate-Pareto-improvement stage  (\crefpart{incentives}{sure-thing} continues to hold even for \cref{improvement}, though), we have decided to nonetheless add this stage as it seems that such examples are delicate enough so that no candidate possesses enough information to plan a successful truncation manipulation.

\section{Outcome of the Match}\label{epilogue}

This market was run (using \cref{improvement}) for the first time in January 2018. A total number of 35 different \Mechina\ programs (administered by 24 independent institutions) offered a cumulative 1,760 slots. A total of 3,120 candidates competed over these slots, of which 2,580 were interviewed by at least one \Mechina\ and ranked at least one \Mechina. Most \Mechinot\ had detailed preferences with many heavily intersecting population definitions, which varied among the \Mechinot:\footnote{We had to personally meet with representatives of some of the \Mechinot, especially those with more complex preferences and population constraints, to walk them through filling out the spreadsheet indicating their preferences. In that sense it seemed that we stretched the cognitive complexity of our spreadsheet-based preference-specification language more or less as far as we could. Such personal meetings became rarer and less crucial in subsequent years, as the different \Mechinot, as well as the Joint Council of \Mechinot, became accustomed to, and more experienced with, our preference-specification language.}
\begin{itemize}
\item
Some population definitions were aimed at balancing the \ofMechinot\ groups. Such were populations reflecting the gender of the candidate, their level of religiousness, and their high school or the geographic location of their home town (many \Mechinot\ looked for balance in the latter, though some \Mechinot\ wanted to make sure that candidates from nearby underprivileged communities have the opportunity to join). Even for common themes such as religiousness, different \Mechinot\ had different granularity, and also classified individual candidates differently than other \Mechinot.
\item
Some population definitions were geared toward affirmative action. Such included minority groups as well as certain groups within the Israeli socioeconomic periphery.
\item
Some population definitions were such that only very few \Mechinot\ or a only single \Mechina\ specifically cared about. These included very specific subgroups within Israeli society, as well as PMA-specific ad-hoc groups of candidates.
\end{itemize}

The incentive properties of \cref{incentives} were explained to the candidates before ranking, using a video that we prepared,\footnote{See timestamps 1:17--3:04 in \url{https://www.youtube.com/watch?v=xt4B2Xu3FvE} . This video also emphasizes other aspects of the market design, e.g., that a \Mechina\ may not pressure a candidate into ranking it high in exchange for a promise to rank the candidate high in return, and that candidates are relieved of any such promises that they gave and are free to rank as they wish \cite[see][]{rn2009}. We included this information in the video following several conversations with candidates. It seems that the video was effective from this angle as well, as this latter problem did not seem to occur in subsequent years.} and in a survey conducted following the match, $92\%$ of candidates reported that they indeed ranked truthfully.\footnote{Of course, we would not have necessarily expected $100\%$ of candidates to be truthful even if the mechanism were precisely strategyproof, see \cite{hmrs2017} and references therein.}
Our  system  matched 1,607 candidates (to $91\%$ of available slots), $85\%$ of whom were matched to their top choice.\footnote{Furthermore, many of the remaining slots were quickly filled in a distributed manner following the run of the algorithm.}

The candidate-Pareto-improvement stage cut the number of blocking pairs of the final match by a factor of $\mbox{$\sim$}1.5$ by resolving four blocking pairs: candidates who would have been unassigned were it not for this stage were assigned to \Mechinot\ that previously rejected them. Following this stage, despite the considerable size of the market, only $10$ pairs were blocking. This may be contrasted with the outcome of the Boston mechanism (also known as the Immediate Acceptance algorithm, see \citealp{aprs2005}), which favors efficiency over stability: on our real-world data the Boston Mechanism would have resulted in $147$ blocking pairs, while not matching more candidates.

Out of the~$10$ blocking pairs in our matching, one was only theoretically problematic as it was only
blocking according to our randomized tie-breaking rather than with respect to the real reported (weak) preferences.
The 9 remaining blocking pairs all involved the same \Mechina. The candidates in all 9 of these blocking pairs were in fact ranked by that \Mechina\ below the ``eviction candidates'' according to the individual ranking by the \Mechina. The only reason each of these pairs was nonetheless blocking is due to one population of that \Mechina\ (the same population for all blocking pairs, to which all blocking candidates belonged) being one candidate short of meeting its minimum target. Thus, resolving any one of these blocking pairs would have caused the rest to no longer block. Moreover, since the minimum target of that population was somewhat of a soft target (``at least $20$ slots for a certain type of geographic population,'' of which $19$ were filled), it is very likely that this \Mechina\ would have decided to not resolve any of these blocking pairs even if it could have.

Finally, we issued the Joint Council of \Mechinot\ a detailed report, to help them answer specific inquiries from \Mechinot\ about specific individual candidates (recall \cref{explainable}), explaining why each candidate that a \Mechina\ ranked above any candidate that it got did not end up being assigned to it (options were because they did not rank that \Mechina, because they were assigned to a \Mechina\ that they preferred, or because adding them would have violated a maximum quota). For the post-matching distributed market (to fill the remaining slots), we gave the \Mechinot\ access to a website that allowed them to check for each candidate whether they are assigned (in which case they are not to be solicited) or not (in which case they are ``fair game''), and allowed candidates to opt-in to a list of interested-and-unassigned candidates that was circulated to \Mechinot\ that did not fill all of their slots in the centralized match.

Following the satisfaction of virtually all of the \Mechinot\ with the groups built out for them by our system in the 2018 match (including the balances of different populations therein), the Joint Council of \Mechinot\ decided to adopt our system for the foreseeable future. The system has already been run twice more since: in January 2019 and most recently in January 2020 --- details regarding these two subsequent matches carried out by our system following the initial 2018 match are given in the next \lcnamecref{subsequent} --- and is slated to run for the fourth time in January 2021.

\section{Further Changes in Subsequent Years}\label{subsequent}

The success of the 2018 match prompted more \Mechinot\ to participate in subsequent matches. The number of candidates steadily increases as well. \cref{stats} summarizes the key statistics of this market since our first match in 2018.%
\begin{table}[t]\onehalfspacing
\centering
\begin{tabular}{cccccccc}
\multirow{3}{*}{Year} & \Mechinot & \multirow{3}{0.9cm}{\centering Total slots} & Candidates
& \multirow{3}{*}{Matches} & Matched & (``Real'') & Unique \\
& (independent & & (submitted & & to their & blocking & blocking \\
& institutions) & & preferences) & & top choice & pairs & \Mechinot \\
\hline
2018 & 35 (24) & 1,760 & $>$3,120\textsuperscript{a} (2,580) & 1,607 & 85\% & 9 & 1\\
2019 & 39 (27) & 2,067 & \phantom{$>$}5,800\phantom{\textsuperscript{a}} (2,842) & 1,801 & 86\% & 9 & 1\\
2020 & 42 (30) & 2,329 & \phantom{$>$}6,118\phantom{\textsuperscript{a}} (3,117) & 1,937 & 84\% & 5 & 1\\
\hline
\end{tabular}

\vspace{.5em}

\begin{minipage}{0.95\textwidth}
\footnotesize\textsuperscript{a}In 2018 the matching web site went live during the interview season, after some candidates were already rejected without interviews by all PMAs to which they applied, so such candidates did not register to the site.
\end{minipage}
\caption{Year-by-year summary statistics of the market: 2018--2020.
}\label{stats}
\end{table}
While the system has remained largely the same since we designed it in 2018, in this \lcnamecref{subsequent} we briefly survey the modifications that we made to the system following its first run in January 2018, for its second run in January 2019.\footnote{We restrict our attention to semantic and algorithmic modifications. The only other modification for the 2019 match was as follows. One of the pain points for \Mechinot\ in the 2018 match in terms of the preference-specification language was populating multiple-valued school and geographical populations (see \cref{multi-valued-populations}). Since belonging to these populations is an objective fact, and since populating them is laborious and prone to mistakes, for the 2019 match onward, we also auto-populated these populations, in the sense that \Mechinot\ could specify maximum quotas and minimum targets for such populations without specifying which candidate belongs to which such population. We also provided a manual override for \Mechinot\ who wanted, e.g., at most 5 students from any one city or town, however allowed up to, say, 7 candidates from one of the two or three largest cities. This manual override required a manual action by us in the 2019 match, but was exposed to the \Mechinot\ via their UI for the 2020 match.}\textsuperscript{,}\footnote{Further modifications to the 2019 system for the third match in January 2020 were not semantic or algorithmic, but rather replaced the Excel spreadsheet UI with a sleeker (yet equivalent in expressiveness) web UI.}

\subsection{Reducing Miscommunication and Misspecifications:\texorpdfstring{\\}{ }Two Further Priority Levels}

One of the prominent criticisms of our 2018 match was by candidates who complained about not being assigned to a certain \Mechina\ whose staff told them they were ranked ``first'' by that \Mechina. In fact, in a survey conducted following the 2018 match, 10\% of the candidates who were not matched to any \Mechina\ claimed that the \Mechina\ that they ranked highest promised them that they would be assigned to it. As it turns out, in many cases such a candidate was in fact assigned rank~``1'' in the Excel spreadsheet by that \Mechina, however was not assigned to that \Mechina\ either because of maximum quota violations that would have occurred because of such an assignment, or since after filling up minimum targets, there were fewer remaining slots at that \Mechina\ than remaining candidates that this \Mechina\ ranked ``1'' in its Excel spreadsheet. This showcased two issues: first, that some \Mechinot\ may not have fully grasped some of the intricacies of the mechanism, and second, that a vague message from a \Mechina\ along the lines of ``we ranked you first/highest'' was confusing at best, and misleading at worst. Another issue with filling in the Excel spreadsheets that would have created problematic effects due to intricacies of our mechanism if it weren't for our manual intervention was already mentioned on \cpageref{malefemale}: many \Mechinot\ initially put minimum targets on gender populations that were the same as the maximum quotas on these populations. Left unchanged, this would have effectively meant that every single candidate helps fill a minimum target, and hence would have completely eliminated the effect of any minimum target (as a candidate helping fill one minimum target gets the same promotion by \cref{choice} as a candidate helping fill several such targets). The most substantial change that we made for the second run of our system in January 2019 involved mitigating both of the above issues.

On the interface level, we added a special ``0-priority'' ranking, which the \Mechinot\ could use to rank any candidate. Raking a candidate as ``0-priority'' has the meaning that such a candidate would be chosen (if she is interested) even if maximum quotas (of specific populations, or even the maximum capacity for the entire \Mechina) are violated. Another modification was to give lesser promotion specifically to candidates that help meet the minimum target of only a gender population (the system recognizes gender populations as such by their names). The full details of how we have been interpreting preference reports as choice functions starting in the 2019 match, with both of these changes, are given in \cref{choice-2019}\begin{algorithm}[t!]
\SetKwInOut{Input}{input}
\SetKwInOut{Output}{output}
\Input{Set $C$ of candidates to choose from}
\Output{Set $\hat{C}$ of candidates chosen from $C$}
\tcp{Initialization}
$\hat{C}\leftarrow\emptyset$\tcp*{Chosen candidates}
\tcp{First pass:\ accept all "0-priority" candidates}
\ForEach{candidate $c\in C$ who is ranked as ``0-priority''}{
$\hat{C}\leftarrow\hat{C}\cup\{c\}$\;
}
\tcp{Second pass:\ give higher priority to candidates who help reach minimum targets of non-gender populations}
\ForEach{candidate $c\in C\setminus\hat{C}$, from highest-ranked to lowest-ranked}{
\If{$c$ is in some non-gender population whose minimum target is not met by $\hat{C}$}{
\If{adding $c$ to $\hat{C}$ does not violate any maximum quota}{
$\hat{C}\leftarrow\hat{C}\cup\{c\}$\;
}
}
}
\tcp{Third pass:\ give higher priority to candidates who help reach minimum targets only of gender populations}
\ForEach{candidate $c\in C\setminus\hat{C}$, from highest-ranked to lowest-ranked}{
\If{$c$ is in some gender population whose minimum target is not met by $\hat{C}$}{
\If{adding $c$ to $\hat{C}$ does not violate any maximum quota}{
$\hat{C}\leftarrow\hat{C}\cup\{c\}$\;
}
}
}
\tcp{Fourth pass:\ all other candidates}
\ForEach{candidate $c\in C\setminus\hat{C}$, from highest-ranked to lowest-ranked}{
\If{adding $c$ to $\hat{C}$ does not violate any maximum quota}{
$\hat{C}\leftarrow\hat{C}\cup\{c\}$\;
}
}
\caption{The choice function of a \Mechina\ from the 2019 match onward.}\label{choice-2019}
\end{algorithm}.

We complemented this addition of ``0-priority'' ranking with an agreement of all \Mechinot\ to increase transparency and clarity by starting any communication with candidates regarding the candidates' ranking with one of three standard messages:
\begin{enumerate}
\item
A message that a \Mechina\ is allowed to send to any candidate, saying that the candidate was found suitable for that \Mechina, but due to the large number of candidates, acceptance cannot be guaranteed at present, and will be determined only during the centralized matching and depends on numbers of candidates and quotas. This message also continues to say that the \Mechina\ would be thrilled if the candidate ranked it high, however also reminded that the Joint Council of \Mechinot\ recommends that each candidate consider multiple possible \Mechinot.
\item
A message that a \Mechina\ is allowed to send \emph{only to candidates ranked by it as \mbox{``priority-0''}}, saying that the candidate was found especially suitable for that \Mechina, and that a slot will be reserved for that candidate (unless the candidate ends up assigned to a \Mechina\ that it ranked higher). This message also emphasizes that no commitment was required by the candidate.
\item
A message that a \Mechina\ is allowed to send only to candidates who are not ranked by it, saying that due to the large number of candidates this year, the candidate will not be considered any further by that \Mechina. This message also continues to say that it does not imply that the candidate is unsuitable for any \Mechina, and that the candidate is encouraged to consider other \Mechinot.
\end{enumerate}

The standardization of initial written responses, together with the \Mechinot\ internalizing that ranking ``priority-1'' does not guarantee acceptance in certain cases, yielded a significant improvement in candidate feedback and satisfaction starting in the 2019 match.

\subsection{A Further Type of Pareto-Improvement Stage}

Manually examining the blocking pairs that resulted from running the 2018 algorithm (\cref{improvement}) on the real-world 2019 data, we noticed a new type of blocking pairs that could be fairly easily resolved without triggering severe ``chain reactions.'' Such a blocking pair involves a \Mechina\ $m$ and an unassigned candidate $c$ that ranks $m$, such that $m$ prefers to replace a candidate $d$ that was assigned by accepting $c$ instead, as long as $d$ has no more options once rejected from $m$, in the strict meaning that every \Mechina\ that $d$ ranks below $m$ does not rank $d$ at all. Such a blocking pair can be easily resolved: simply assign $c$ to $m$ instead of $d$, and  leave $d$ unassigned.\footnote{Note that while neither $d$ nor can form any new blocking pairs following this resolution, in fact $m$ might form new blocking pairs, for example if $d$ belongs to different populations than $c$, however trial runs and real-world data show this to be fairly rare.} Observing that our preliminary match had a number of such blocking pairs, we heuristically tweaked the final stage of our algorithm to resolve such pairs, resulting in the final algorithm that we ran in 2019.\footnote{Technically, after computing a matching via DA, we alternated between a candidate-Pareto-improvement stage as in \cref{improvement} and a \Mechina-Pareto-improvement stage as described above.
}

As with the candidate-Pareto-improvement stage (from \cref{improvement}), even adding this \Mechina-Pareto-improvement stage alone would have broken \crefpart{incentives}{truncation} (in fact, this new type of Pareto-improvement stage also creates incentives to extend one's list under certain circumstances, and not only to truncate it under other circumstances), but adding even both types of Pareto-improvement stages does not break \crefpart{incentives}{sure-thing}. As with the candidate-Pareto-improvement stage, we have similarly decided to nonetheless heuristically add the \Mechina-Pareto-improvement stage as it seems that the information needed to plan a successful truncation (or other) manipulation with this stage is still far beyond the reach of any candidate. The addition of this Pareto-improvement stage cut the number of blocking pairs of the final 2019 match (computed on real data) by a factor of $\mbox{$\sim$}2$ compared to having only the candidate-Pareto-improvement stage as in \cref{improvement}.

While in each of the 2018 and 2019 matches we introduced a new type of Pareto-improvement stage based on the blocking pairs that manifested in that year's specific data, we remark that these Pareto-improvement stages have proven to be more robust and useful than one may have expected given the circumstances of their initial introductions. Indeed, for the 2020 match (computed on real data), the combination of these two Pareto-improvement stages (which we used unaltered for~2020) cut the number of real blocking pairs by a factor of $\mbox{$\sim$}10$, from $53$ to $5$.

\vspace{.5em}

Overall, as our original design and the above modifications demonstrate, constructing a successful matching system has as much human engineering in it, if not more, as it does theory. Moreover, being constrained by the requirement to build an actual matching system necessitated that we shift our focus from more traditional theoretical results of the form ``under so and so conditions, a stable matching exists, and otherwise it is infeasible to determine whether one even exists'' to a more ``practical counterpart'' to that approach: we designed a matching system that under the same conditions finds a stable matching, and otherwise finds a matching for which we heuristically aimed to be (and which empirically turned out to be in all years so far) ``almost stable.'' Similarly, while we had to sacrifice strategyproofness (at least theoretically), we however were able to recover strategic guarantees in proving that our mechanism is ``strategyproof enough'' in a precise novel sense, of being immune to the manipulations that applicants considered in practice. Taking theory to practice indeed involved many challenges, but also yielded answers to theoretical questions that we would never have thought to ask were it not for our involvement in this practical project.

\bibliographystyle{abbrvnat}
\bibliography{paper}

\clearpage

\renewcommand{\topfraction}{.9}
\renewcommand{\textfraction}{.1}

\appendix

\section{Proofs}\label{analysis}

\begin{proof}[Proof of \cref{laminar-chains}]
We will show that for the assumed population structure, any choice function induced by \cref{choice-prefer-more} is substitutable and satisfies irrelevance of rejected contracts and the law of aggregate demand. By the analysis of \cite{hm2005} and \cite{as2013}, this will imply the theorem. More specifically, we will show that for every candidate set $C$ and for every $c\in C$, if we denote by $\hat{C}$ the set chosen from $C$ by a given \Mechina\ and by $\hat{C}'$ the set chosen from $C\setminus\{c\}$ by the same \Mechina\ (with the same preferences), then if $c\notin\hat{C}$ then $\hat{C}'=\hat{C}$ (irrelevance of rejected contracts), and otherwise either $\hat{C}'=\hat{C}\setminus\{c\}$ or $\hat{C}'=(\hat{C}\setminus\{c\})\cup\{d\}$ for some $d\in C$. This will in particular imply that $\hat{C}\setminus\{c\}\subseteq\hat{C}'$ (substitutability) and that $|\hat{C}'|\le|\hat{C}|$ (law of aggregate demand).

In this proof we will trace the invocation of \cref{choice-prefer-more}, in parallel both for the input $C$, which we will call the ``original run'' of the \lcnamecref{choice-prefer-more}, and for the input $C\setminus\{c\}$, which we will call the ``new run'' of the \lcnamecref{choice-prefer-more}. To be more specific, we consider the equivalent implementation of \cref{choice-prefer-more} that is given in \cref{choice-prefer-more-parallel}\begin{algorithm}[t]
\SetKwInOut{Input}{input}
\SetKwInOut{Output}{output}
\Input{Set $C$ of candidates to choose from}
\Output{Set $\hat{C}$ of candidates chosen from $C$}
\tcp{Initialization}
$\hat{C}\leftarrow\emptyset$\tcp*{Chosen candidates}
$n\leftarrow$ the number of populations with a (positive) minimum target\;
\For{$i\gets n$ \KwTo $0$ \KwBy descending order}{
\ForEach{candidate $c\in C$, from highest-ranked to lowest-ranked}{
\If{$c$ is in $n$ populations whose minimum target is not met by $\hat{C}$}{
\If{adding $c$ to $\hat{C}$ does not violate any maximum quota}{
$\hat{C}\leftarrow\hat{C}\cup\{c\}$\;
}
}
}
}
\caption{Equivalent implementation of \cref{choice-prefer-more} (along the lines of the implementation of \cref{choice}) that is traced, in parallel for two input candidate sets, in the proof of \cref{laminar-chains}.}\label{choice-prefer-more-parallel}
\end{algorithm}, and trace the corresponding iterations of the double-loop in both runs. We will refer to a single iteration of the inner loop as a \emph{step} of the \lcnamecref{choice-prefer-more-parallel}.
We observe that a property of \cref{choice-prefer-more} (and in fact, of \cref{choice} as well) is that at any point, the set of candidates chosen so far (along with the reported preferences) effectively defines an order on all not-yet-considered candidates, according to which the next candidate to be considered is chosen. Therefore, if candidate $c$ is not chosen at any point in the original run, then these two runs have the same candidates considered at the same steps, and hence the same candidates chosen, and at the same steps, throughout the run --- this is true even in the absence of any structural assumptions on the populations. Otherwise, by the same reasoning (and still without using any structural assumption), if candidate $c$ is chosen in the original run, then that must be the first difference between the runs: at that point $c$ is chosen in the original run but is not chosen in the new run, as it is not considered in the latter, since it is not in the set of candidates to choose from.
From that point onward, as long as the same candidates are chosen in both runs and in the same steps, for any population the number of chosen candidates from this population in the new run is less than or equal to the same number in the old run. Therefore, either both runs accept the exact same candidates from that point, in which case $\hat{C}'=\hat{C}\setminus\{c\}$, or the chronologically next difference in chosen candidates between both runs is that some candidate $d$ is chosen in the new run but not in the corresponding step in the original run. We will continue the proof by induction over the number of candidates chosen (say, in the new run) until the step in which $d$ is chosen, where the induction is in decreasing order (so formally, the induction is over $|C|$ minus this number).

Since in the original run $d$ is not chosen at the same step in which she is chosen in the new run, this is either because she violates the maximum quota of a population to which $c$ belongs, or she does not receive as high a promotion in the new run because adding $c$ satisfies a minimum target for some population to which $d$ belongs. Let us first consider the former case. In this case, adding~$d$ in the new run causes the number of chosen candidates from that maximum-quota population $P$ to equal its maximum quota (since adding $c$ instead would have done so). So, after the step in which $d$ is added in the new run, in neither run will candidates from population $P$ be chosen. We claim therefore that the same candidates are chosen at the same steps in both runs from that point onward. This can be seen by induction: for any candidate not from $P$ (as no more candidates from $P$ are chosen) considered at any time after $d$ is chosen in the new run, we claim that for each population $P'$ to which this candidate belongs, the same number of candidates from $P'$ are chosen in both runs by that time. Indeed, since this candidate is not from $P$, by laminarity either population $P'$ is disjoint from $P$ or it contains $P$, so, since by the induction hypothesis the only difference between the sets of chosen candidates in both runs by that time is that $c$ is chosen in the original run while $d$ is chosen in the new run, then the same number of candidates from $P'$ have been chosen in either run, so the candidate at that time is either chosen in both runs or chosen in neither run. Therefore, in this case $\hat{C}'=(\hat{C}\setminus\{c\})\cup\{d\}$ and we are done.\footnote{As mentioned in \cref{algorithm-barriers}, the proof of \cref{laminar-chains} up until this point, when applied in the absence of minimum targets/quotas (in which case \cref{choice} and \cref{choice-prefer-more} coincide with each other and with the choice function of \citealp{huang2010}), constitutes not only a full concise alternative proof of the theorem of \cite{huang2010} that in the absence of any minimum quotas a stable matching exists, but also a proof that the choice function is substitutable and satisfies irrelevance of rejected contracts and the law of aggregate demand, which implies many additional desirable properties \citep{roth1984,hm2005,as2013} including that DA is strategyproof.} In the rest of this proof we will therefore consider the case in which $d$ is not chosen in the original run at the same step in which she is chosen in the new run because adding $c$ satisfies a minimum target for some population to which $d$ belongs. In particular, we note that in this case $c$ and $d$ share a minimum-target population.

If $\hat{C}'=(\hat{C}\setminus\{c\})\cup\{d\}$, then we are done. So, assume that this is not the case, and we will look at the next chronological difference (which by this assumption must exist) in acceptances after the step in which in the new run $d$ is chosen. We reason by cases based on whether this difference is for a candidate $e$ to be chosen in the original run but not in the corresponding step in the new run, or for a candidate $e$ to be chosen in the new run but not in the corresponding step in the original run.

\textbf{Consider first the case in which a candidate $\boldsymbol e$ is chosen in the original run but not (at the corresponding step at least) in the new run.} Assume for contradiction that $e\ne d$. Since in the new run $e$ is not chosen in the step in which it is chosen in the original run, this is either because it violates the maximum quota of a population to which $d$ but not $c$ belongs, or it did not receive as high a promotion in the new run because adding $d$ satisfies more minimum targets for populations to which $e$ belongs than adding $c$ does. Either way, $e$ shares a population with $d$ to which $c$ does not belong. Therefore, by laminarity of the population structure, any population that contains $c$ and $d$ also contains $e$, and any population that contains $c$ and $e$ also contains $d$. So, the minimum-target population that we have shown that $c$ and $d$ share also contains $e$. Therefore, $c$, $d$, and $e$ share a minimum-target population, and so by the assumption on the structure of minimum-target populations, these three candidates belong to the same minimum-target population chain.

We claim that in any step from the time at which $c$ is chosen in the original run and until the end of each run separately, $d$ and $e$ are in the exact same number of populations whose minimum targets are at that step not yet met. Indeed, if at any step $d$ (resp.\ $e$) were in more such populations than $e$ (resp.\ $d$), then since these two candidates belong to the same minimum-target population chain, this means that $d$ (resp.\ $e$) belongs to a deeper population in that chain than $e$ (resp.\ $d$), and that the minimum target of this ``deeper'' population is not yet met at that step.
Since any population that contains $c$ and $d$ also contains $e$ (resp.\ since any population that contains $c$ and $e$ also contains $d$), and since all three candidates belong to that minimum-target population chain, this means that $d$ (resp.\ $e$) belongs to a deeper population in that chain than also $c$, and that the minimum target of this ``deeper'' population is not yet met at that step. Since the minimum target of this ``deeper'' population is not yet met at that step, it is also not met at any previous step. In particular, it is not met just before $c$ is chosen in the original run (as the runs are identical before $c$ is added),
and therefore in the beginning of that step $d$ (resp.\ $e$) is in more populations whose minimum target is not yet met than $c$ is, and so $d$ (resp.\ $e$) should have been chosen even before $c$, as choosing it would not have violated any maximum quotas, since $d$ is chosen later in the new run and added to a superset of the same chosen set of candidates without violating any maximum quotas (resp.\ since $e$ is later chosen in the original run).
This is a contradiction, and so indeed in any step from the time at which $c$ is chosen in the original run and until the end of each run separately, $d$ and $e$ are in the exact same number of populations whose minimum targets are at that step not yet met.

Since adding $e$ does not violate any maximum quotas when it is chosen in the original run, adding it earlier in the original run, just before $d$ is chosen in the new run, would not have violated any maximum quotas either, and so, since in the corresponding step in the new run the set of already-chosen agents is smaller, adding $e$ in the new run at that step would not have violated any maximum quotas either. Since at that step $d$ and $e$ are in the exact same number of populations whose minimum targets are not yet met, and since $d$ is chosen at that step and $e$ is not added before even though both additions are feasible in terms of not violating any maximum quotas, we conclude that the given \Mechina\ prefers $d$ to $e$.

Since $e$ is chosen in the original run when $d$ is not (yet) chosen, since at that step $d$ and $e$ are in the exact same set of populations whose minimum targets are not yet met, and since the given \Mechina\ prefers $d$ to $e$, this means that choosing $d$ at that step would have violated a maximum quota for a population to which $e$ does not belong. Since any population to which $c$ and $d$ belong also contains $e$, this means that neither does $c$ belong to this maximum-quota population, and so choosing $d$ at that step (of the original run) would have violated the same maximum quota even if the set of already-chosen candidates at the time were the same set only with $c$ removed, however this set with $c$ removed is precisely the set of already-chosen candidates in the corresponding step in the new run, and $d$ is chosen in that step in that run, so no maximum quotas are violated by adding $d$ to this set --- a contradiction. So, $e=d$.

Immediately following the step in which $e=d$ is chosen in the original run, the sets of chosen candidates in both runs differ only by $c$ being only chosen in the first run. At this point we are either done if no more differences in acceptances between the two runs occur, or (by the same reasoning as in the beginning of this proof, since for any population, the number of chosen candidates from this population in the new run is less than or equal to the same number in the old run), the next chronological difference is for a candidate $d'$ to be chosen in the new run but not in the old run. In this case by the induction hypothesis we are also done.

\textbf{We now consider the case in which a candidate $\boldsymbol e$ is chosen in the new run but not (at that point at least) in the original run.} We will complete the proof by showing that this case leads to a contradiction. Since in the original run $e$ is not chosen in the step in which it is chosen in the new run, this is either because it violates the maximum quota of a population to which $c$ but not $d$ belongs, or it does not receive as high a promotion in the original run because adding $c$ satisfies more minimum targets for populations to which $e$ belongs than adding $d$ does. Either way, $e$ shares a population with $c$ to which $d$ does not belong. Therefore, by laminarity of the population structure, any population that contains $c$ and $d$ also contains $e$, and any population that contains $d$ and $e$ also contains $c$.  So, the minimum-target population that we have shown that $c$ and $d$ share also contains $e$. Therefore, $c$, $d$, and $e$ share a minimum-target population, and so by the assumption on the structure of minimum-target populations, these three candidates belong to the same minimum-target population chain.

We claim that in any step from the time at which $c$ is chosen in the original run and until the end of each run separately, $e$ is in at least as many populations whose minimum targets are at that step not yet met as $d$ is. Indeed, if at any step $d$ were in more such populations than $e$, then since they belong to the same minimum-target population chain, this means that $d$ belongs to a deeper population in that chain than $e$, and that the minimum target of this ``deeper'' population is not yet met at that step. Since any population that contains $c$ and $d$ also contains $e$, and since all three candidates belong to that minimum-target population chain, this means that $d$ belongs to a deeper population in that chain than also $c$, and that the minimum target of this ``deeper'' population is not yet met at that step. Since the minimum target of this ``deeper'' population is not met at that step, it is also not met at any previous step. In particular, it is not met just before $c$ is chosen in the original run (as the runs are identical before $c$ is added),
and therefore in the beginning of that step $d$ is in more populations whose minimum target is not yet met than $c$ is, and so $d$ should have been chosen even before $c$, as choosing it would not have violated any maximum quotas, since $d$ is chosen later in the new run and added to a superset of the same chosen set of candidates without violating any maximum quotas. This is a contradiction, and so indeed in any step from the time at which $c$ is chosen in the original run and until the end of each run separately, $e$ is in at least as many populations whose minimum targets are at that step not yet met as $d$ is.

Since $e$ is chosen after $d$ in the new run, adding it at any time before $d$ is chosen would not have violated any maximum quotas either. Since at that step $e$ is in at least as many populations whose minimum targets are not yet met as $d$, and since $d$ is chosen at that step and $e$ is not added before even though both additions are feasible in terms of not violating any maximum quotas, we conclude that the given \Mechina\ prefers $d$ to $e$.

Since any population that contains $c$ and $d$ also contains $e$, and any population that contains $d$ and $e$ also contains $c$, the set of populations that contain $c$ and $d$ is precisely the same as the set of populations that contain $d$ and $e$ (as both of these are precisely the same as the set of populations that contain $c$, $d$, and $e$).

Consider the original run in the step in which $e$ is chosen in the new run. Adding $e$ in this step in the new run does not violate any maximum quotas. Therefore, adding $d$ in the same step in the original run would not have violated any maximum quotas, because any maximum quotas violated by this would have still been violated even if the set of chosen candidates at the time were the same set with $c$ replaced by $e$ (since the populations that contain $c$ and $d$ are the same as the populations that contain $d$ and $e$), however in the corresponding step in the new run, this set with the addition of $d$ is precisely the set of chosen candidates once $e$ is chosen, and so no maximum quotas are violated by it.

Still considering the step in the which $e$ is chosen in the new run, we now claim that the number of populations whose minimum targets are at that step in the original run not met and to which $d$ belongs, is the same as the number of populations whose minimum targets are at that step in the new run not met and to which $e$ belongs. To see this, we first note that any population that counts toward the former number must have $c$ in addition to $d$ in it (otherwise $d$ is deeper than $c$ in the chain of minimum-target populations to which they both belong, and furthermore the minimum target of this population is not yet met also just before $c$ is chosen in the original run, and so $d$ should have been chosen before that in that run), and any population that counts toward the latter must have $d$ in addition to $e$ it (otherwise $e$ is deeper than $d$ in the chain of minimum-target populations to which they both belong, and furthermore the minimum target of this population is not yet met also just before $d$ is chosen in the new run, and so $e$ should have been chosen before that in that run). Therefore, since the set of populations that contain $c$ and $d$ is precisely the same as the set of populations that contain $d$ and $e$, and both of these are precisely the same as the set of populations that contain $c$, $d$, and $e$, only populations that contain $c$, $d$, and $e$ count toward any of these two numbers. Now, each of these populations can be tested to count toward the former number by counting the number of chosen candidates at that step (in the original run), except for $c$, that are in that population, adding one (since $c$ is in that population), and checking if this satisfies its minimum target. Similarly, each of these populations can be tested to count toward the latter number by counting the number of chosen candidates at that step (in the new run), except for $d$ (this is exactly the same as the chosen candidates at that step in the original run except for $c$!) that are in that population, adding one (since $d$ is in that population), and checking if this satisfies its minimum target. Therefore, each of these populations either counts toward both of these numbers or toward none of them, and so these numbers are indeed the same.

To sum up, in the step of the original run that corresponds to the step of the new run in which $e$ is chosen, adding $d$ would not have violated any maximum quotas, and the number of populations with unmet minimum targets at that step in the original run to which $d$ belongs is such that $d$ at that step (and before) passes the minimum-target conditional of the iteration of the outer loop of \cref{choice-prefer-more-parallel} that contains that step. Since the given \Mechina\ prefers $d$ to $e$, this means that $d$ should have in fact already been chosen in the original run at that point --- a contradiction to the assumption that the next difference between the two runs, following the addition of $d$ in the new run, is the addition of $e$ in the new run.
\end{proof}

\begin{proof}[Proof of \cref{impossibility}]
For stability,
consider a market with three candidates $c$, $d$, $e$, and two \Mechinot\ $m$ and $m'$. The preferences of the candidates are:
\[ c: m' \succ m \qquad d: m \succ m' \qquad e: m \succ m' \]
The candidate ranking of \Mechina~$m$ is: $c\succ d\succ e$. The populations that~$m$ considers are \{c,d,e\} with minimum target $1$, \{d,e\} with maximum quota $1$, and \{e\} with minimum target $1$.
The candidate ranking of \Mechina~$m'$ is: $e\succ c\succ d$. There are no populations for~$m'$, and the overall maximum quota of~$m'$ is $1$.

To see that no stable matching (with respect to the choice functions induced by \cref{choice} given the above preference reports) exists in this market, we reason by cases based on whether or not candidate~$c$ is assigned to \Mechina~$m$ in a proposed matching. If $c$ is assigned to $m$, then since $c$ prefers $m'$ over $m$, stability dictates that $m'$ prefer the candidate assigned to it over $c$. So, $e$ is assigned to $m'$. But then we have that candidate~$e$ blocks with \Mechina~$m$, since $e$ prefers $m$ over $m'$, and $m$ chooses the pair $\{c,e\}$ from any candidate set that contains $\{c,e\}$ (and $c$ is assigned to $m$ by assumption).

If $c$ is not assigned to $m$, then since $d$ ranks $m$ highest and since $m$ chooses $\{d\}$ from any candidate set that contains $d$ but does not contain $c$, by stability $d$ is alone assigned to $m$. Therefore, $e$ is not assigned to $m$, and so by stability $e$ is assigned to her next choice $m'$, since $m'$ ranks her highest. Therefore, $c$ is unassigned, but then we have that $c$ blocks with $m$, since $m$ chooses $\{c,d\}$ from $\{c,d\}$ (and $\{d\}$ is assigned to $m$).

For strategyproofness, consider a market with five candidates $c$, $d$, $e$, $f$, $g$, and three \Mechinot\ $m$, $m'$, $m''$. The preferences of the candidates are (these are \Mechinot\ ranked highest by each candidate; other \Mechinot\ can added to these rankings in lower places):

\[ c:m''\succ m \qquad d:m \qquad e:m\succ m' \qquad f:m'' \qquad g:m' \]

The candidate ranking and populations of \Mechina\ $m$ are as in the stability counterexample above. Each other \Mechina\ has an overall maximum quota of $1$ and no populations. \Mechina\ $m'$ ranks $g$ first, and then all other candidates in some order, and \Mechina\ $m''$ ranks $f$ first, and then all other candidates in some order.

If all candidates report their preferences truthfully, a run of the algorithm is as follows: the first proposals are by $d$ and $e$ to $m$, by $g$ to $m'$, and by $c$ and $f$ to $m''$. The algorithm proceeds by rejecting $e$ from $m$, rejecting $c$ from $m''$, and rejecting $e$ from $m'$, resulting in the final matching of $c$ and $d$ to $m$, of $g$ to $m'$, of $f$ to $m''$, and of $e$ to no \Mechina.

We will show that $e$ has incentive to misreport its preferences as $m'\succ m$. Indeed, a run of the algorithm in this case is as follows: the first proposals are by $d$ to $m$, by $e$ and $g$ to $m'$, and by $c$ and $f$ to $m''$. Then algorithm proceeds by rejecting $c$ from $m''$ (as in the original run) and rejecting $e$ from $m'$. The crux of this counterexample is that in this run when $e$ proposes to $m$, also $c$ proposes to it, causing $d$ (rather than $e$) to be rejected from $m$, and resulting in the final matching of $c$ and $e$ to $m$, of $g$ to $m'$, of $f$ to $m''$, and of $d$ to no \Mechina --- a matching strictly preferred by $e$ according to $e$'s true preferences.
\end{proof}

\begin{proof}[Proof of \cref{incentives}]
For \cref{incentives-truncation}, note that a run of candidate-proposing DA with the original non-truncated preference list for the given candidate would be identical to the run with the truncated preference list, as the candidate would never be rejected from the given \Mechina\ in the former run, and so any \Mechinot\ ranked lower than that \Mechina\ would never be accessed.

For \cref{incentives-sure-thing}, note that in candidate-proposing DA, the given candidate will be assigned to a \Mechina\ that she placed lower than the given \Mechina\ on her preference list only if the given \Mechina\ rejects her in some round (that is, does not choose her in some round in which this candidate applies to it). As it is assumed that the given \Mechina\ always chooses this candidate from any set of candidates that contains it, this cannot be, and so the candidate is either assigned to the given \Mechina\ or to one that she has placed higher on her preference list.
\end{proof}

\end{document}